\documentclass{ws-procs9x6}            % default, citations in superscript
\begin{document}
\title{Multi-Soliton scattering of the Anti-Self-Dual Yang-Mills Equations in 4-dimensional split signature}

\author{Shan-Chi Huang\footnote{This work is supported by Grant-in-Aid for Scientific Research (18K03274) and the scholarship of Japan-Taiwan Exchange Association.}}

\address{Graduate School of Mathematics, Nagoya University,\\
Nagoya, 464-8602, Japan\\
$^*$E-mail: x18003x@math.nagoya-u.ac.jp}

%\author{A. N. Author}

%\address{Group, Laboratory, Street,\\
%City, State ZIP/Zone, Country\\
%E-mail: an\_author@laboratory.com}

\begin{abstract}
We construct the ASDYM 1-solitons and multi-solitons for split signature and interpret them as soliton walls. 
We show that the gauge group is $\mathrm{G=SU(2)}$ for the entire intersecting soliton walls, and $\mathrm{SU(3)}$ for each soliton walls in the asymptotic region. 
This is joint research partially with Masashi Hamanaka, Claire R. Gilson, and Jonathan Nimmo.
\end{abstract}


\bodymatter

\section{Introduction}
The anti-self-dual Yang-Mills (ASDYM) system is notable as an exactly solvable model and play an indispensable role in many different fields of theoretical and mathematical physics, such as quantum field theory, twistor theory \cite{MaWo}, and integrable systems \cite{MaWo}. 
Especially the ASDYM equations reveal the essence of the lower-dimensional integrable systems because almost all of the soliton equations are the dimensionally reduced equations of ASDYM according to the Ward conjecture \cite{Ward}. 
Among these lower-dimensional soliton equations, dimensionally reduced from  the 4-dimensional split signature ($+$, $+$, $-$, $-$) (or called the Ultrahyperbolic \cite{MaWo} spacetime) are the majority. 
On the other hand, 
the ASDYM equations in the 4-dimensional split signature is the EOM of effective action for $\mathrm{N=2}$ open string theories \cite{OoVa} and seem to have good compatibility with the lower-dimensional integrable systems. 
Although most physicists didn't pay too much attention on $\mathrm{N=2}$ strings and there has been little progress in this direction for a long time, we still believe that the connections between the lower-dimensional integrable systems and $\mathrm{N=2}$ open string theories are worth studying and not just for seeking the physical reality.

In (1+1)-dimensional integrable systems, the KdV equation
%\begin{eqnarray}
$-4u_{t}+6uu_{x}+u_{xxx}=0$
%\end{eqnarray}
might be the most notable one among the exactly solvable models and the typical KdV 1-soliton is in the form of 
\begin{eqnarray}
\label{KdV 1-soliton}
u(x,t)
:=2\frac{\partial^{2}}{\partial x^{2}}(\log\mbox{cosh}X)
=2\kappa^{2}\mbox{sech}^{2}X,~~X:=\kappa x+\kappa^{3}t + \delta
\end{eqnarray}
which is a travelling wave solution with a constant velocity $-\kappa^2$ and a constant amplitude $2\kappa^2$, and hence it preserves its shape over time. In addition, $u(x,t)$ is an even function symmetric with respect to $X=0$ and $u, u_x, u_{xx}, u_{xxx} \rightarrow 0$ as $x \rightarrow \pm \infty$. Therefore, the distribution of $u(x,t)$ is localized within a region centered on $X=0$. 
On the other hand, the KdV equation is also known for having infinitely conserved quantities \cite{MiGaKr}, more specifically, 
the mass $\int u \mbox{d}x$, the momentum $\int u^{2} \mbox{d}x$, 
the energy $\int \left[ 2u^3 -(u_{x})^2 \right]\mbox{d}x$, ..., and so on. 
In particular, the energy density 
\begin{eqnarray}
\label{KdV energy density}
2u^3 -(u_{x})^2=16\kappa^{6}\left( 2\mbox{sech}^6X-\mbox{sech}^4X \right), ~~
X:=\kappa x+\kappa^{3}t + \delta
\end{eqnarray}
possesses the behavior of 1-soliton as well.
Inspired by this, we simply consider the pure Yang-Mills action density Tr$F_{\mu\nu}F^{\mu\nu}$ as
an analogue of energy density, and try to figure out whether the 1-solitonic behavior like  \eqref{KdV energy density} exists for the ASDYM equations in 4-dimensions or not. Fortunately, the answer is yes and our result \cite{HaHu1} is 
\begin{eqnarray}
{\mbox{Tr}} F_{\mu \nu}F^{\mu \nu} \propto \left(2{\mbox{sech}}^2 X-3{\mbox{sech}}^4 X\right),
\end{eqnarray} 
where X is a nonhomogeneous linear function of real coordinates $x^{1}$, $x^{2}$, $x^{3}$, $x^{4}$. 
Especially, the gauge group is $\mathrm{G=SU(2)}$\cite{HaHu1} for the split signature and hence it means our ASDYM 1-soliton could be some candidate of physical object in $\mathrm{N=2}$ open string theories.

One more well-known fact is that the KdV multi-soliton 
\begin{eqnarray}
u(x,t)
:=
2\frac{\partial^{2}}{\partial x^{2}}(\log\tau_{n})
\end{eqnarray}
can be represented elegantly as the Wronskian determinant 
\begin{eqnarray}
\label{Wronskian}
&&\tau_{n}:=
\mbox{Wr}(f_1, f_2, ..., f_n)  
:=
\left|
\begin{array}{ccccc}
f_1^{~(0)} & f_2^{~(0)} & \cdots & f_n^{~(0)} \\
f_1^{~(1)} & f_2^{~(1)} & \cdots & f_n^{~(1)} \\
\vdots & \vdots & \ddots & \vdots \\
f_1^{~(n-1)} & f_2^{~(n-1)} & \cdots & f_n^{~(n-1)}
\end{array}
\right|, ~~~~~~~~    \\
&& {\mbox{where}}~ 
f_i^{~(m)}:=\frac{\partial^{m}f_i}{\partial x^{m}},~
f_i:=
\mbox{cosh}X_i,~~
X_i:=\kappa_i x+\kappa_i^{3}t + \delta_i.
\end{eqnarray}
Now a natural question to ask is whether the Wronskian type multi-solitons exist for the ASDYM equations or not. 
If the answer is yes and the gauge group can be verified to be unitary, then what is interpretation for such ASDYM multi-solitons in $\mathrm{N=2}$ string theories.

\section{$J$-matrix formulation of ASDYM and Darboux transformation}

Before going ahead to the main topic, we need to introduce some prior knowledge.
Firstly, the ASDYM equations on various real spaces can be unified into 4-dimensional complex flat spacetime with the metric $ds^2=2(dz d\widetilde z -dw d\widetilde w)$ and we can easily get the split signature ($+$, $+$, $-$, $-$) by imposing some conditions on the complex coordinates, for instance
\begin{eqnarray} 
\label{Reality conidtion_U}
	\left(\begin{array}{cc}
	z & w \\
	\widetilde{w} & \widetilde{z}
	\end{array} \right)
	=
	\frac{1}{\sqrt{2}}
	\left(\begin{array}{cc}
	x^{1}+x^{3} & x^{2}+x^{4} \\
	-(x^{2}-x^{4}) & x^{1}-x^{3} 
	\end{array}\right), ~ x^{1}, x^{2}, x^{3}, x^{4} \in \mathbb{R}. 
\end{eqnarray}
Here we set the gauge group to be $\mathrm{G=GL(N,\mathbb{C})}$ in general.
The complex representation of ASDYM equations are
\begin{eqnarray}
F_{zw}=0,~ F_{\widetilde{z}\widetilde{w}}=0,~F_{z\widetilde{z}} - F_{w\widetilde{w}}=0   
\end{eqnarray}
which can be cast in a gauge independent formulation, called the Yang equation \cite{Yang, BrFaNuYa}
\begin{eqnarray}
\label{Yang equation}
\partial_{\widetilde{z}}[(\partial_{z}J)J^{-1}] - \partial_{\widetilde{z}}[(\partial_{z}J)J^{-1}] =0, 
\end{eqnarray}
where $J$ is an $N \times N$ matrix called Yang's $J$-matrix. An advantage of this formulation is that the anti-self-dual (ASD) gauge fields can be reformulated by the decomposition of  $J=\widetilde{h}^{-1}h$ as 
\begin{eqnarray}
A_{z}\!=\!-(\partial_{z}h)h^{-1}, A_{w}\!=\!-(\partial_{w}h)h^{-1},
A_{\widetilde{z}}\!=\!-(\partial_{\widetilde{z}}\widetilde{h})\widetilde{ h}^{-1},
A_{\widetilde{w}}\!=\!-(\partial_{\widetilde{w}}\widetilde{h})\widetilde{h}^{-1}. ~~~~  
\end{eqnarray}
Sometimes we use a convenient gauge $\widetilde{h}=1$ to simplify the gauge fields as
\begin{eqnarray}
\label{Gauge fields}
A_{z}=-(\partial_{z}J)J^{-1}, ~A_{w}=-(\partial_{w}J)J^{-1},~ 
A_{\widetilde{z}}=0,~
A_{\widetilde{w}}=0. 
\end{eqnarray}
Now we introduce a novel Lax representation \cite{NiGiOh} of ASDYM equations :
\begin{eqnarray}
\label{Linear system_NGO}
\left\{
\begin{array}{c}
L(\phi):= [\partial_{w}-(\partial_{w}J)J^{-1}]\phi-(\partial_{\widetilde{z}}\phi)\zeta = 0 \\
M(\phi):=	
[\partial_{z}-(\partial_{z}J)J^{-1}]\phi-(\partial_{\widetilde{w}}\phi)\zeta = 0   
\end{array}
\right.  ,~~ (\mathrm{G=GL(N,\mathbb{C})})
\end{eqnarray}
where the spectral parameter $\zeta$ is any right-action $N \times N$ constant matrix rather than scalar.
Under the compatible condition $L(M(\phi))-M(L(\phi))=0$,
the ASDYM equations (Yang equation) can be derived from the linear system \eqref{Linear system_NGO}.
Furthermore, if we define a Darboux Transformation \cite{NiGiOh} as
\begin{eqnarray}
\label{Darboux transf}
\widetilde{\phi}=\phi\zeta-\psi\Lambda \psi^{-1}\phi,~~
\widetilde{J}=-\psi\Lambda \psi^{-1}J,   
\end{eqnarray}
where $\phi(\zeta)$ denotes the general solution of \eqref{Linear system_NGO} and we use a new notation $\psi(\Lambda)$ to denote a specified solution of \eqref{Linear system_NGO}.
Then the linear system \eqref{Linear system_NGO} is form invariant under the Darboux transformation \eqref{Darboux transf}. In other words, 
\begin{eqnarray}
\label{Linear system_NGO_2}
\left\{
\begin{array}{c}
\widetilde{L}(\widetilde{\phi}):= [\partial_{w}-(\partial_{w}\widetilde{J})\widetilde{J}^{-1}]\widetilde{\phi}-(\partial_{\widetilde{z}}\widetilde{\phi})\zeta = 0 \\
\widetilde{M}(\widetilde{\phi}):=	
[\partial_{z}-(\partial_{z}\widetilde{J})\widetilde{J}^{-1}]\widetilde{\phi}-(\partial_{\widetilde{w}}\widetilde{\phi})\zeta = 0
\end{array}
\right..  
\end{eqnarray}
Now we can choose a seed solution $J=J_1$ of the Yang equation \eqref{Yang equation} and substitute it into \eqref{Linear system_NGO} to solve a specified solution $\psi_{1}(\Lambda_1)$. 
After 1-iteration of the Darboux transformation \eqref{Darboux transf}, we get a new solution $\widetilde{J}=J_2$. By repeating the same process\cite{SCH}, we can obtain a series of $J$-matrices.

~~~~{\bf \scriptsize{Seed solution}}  \\
$~~~~~~~~~~~ \overbrace{J_1} ~\stackrel{\bf Dar}{\longrightarrow} ~ J_2 ~ \stackrel{\bf Dar}{\longrightarrow} ~
J_3 ~ \stackrel{\bf Dar}{\longrightarrow} ~ J_4 ~ \stackrel{\bf Dar}{\longrightarrow}~ ...~\stackrel{\bf Dar}{\longrightarrow}~J_{n+1} ~ \stackrel{\bf Dar}{\longrightarrow}~...$ \\
and the $J$-matrix $J_{n+1}$ can be written concisely in terms of the Wronskian\cite{GiHaHuNi} type quasideterminant (Cf: \eqref{Wronskian}) :
\begin{eqnarray}
\label{J_n+1}
J_{n+1} 
=\left|
\begin{array}{ccccc}
\psi_{1} & \psi_{2} & \!\! \cdots & \!\! \psi_{n} & 1 \\
\psi_{1}\Lambda_{1} & \psi_{2}\Lambda_{2} & \!\! \cdots & \!\! \psi_{n}\Lambda_{n} & 0 \\
\vdots &  \vdots & \!\! \ddots & \!\! \vdots & \vdots  \\
%\psi_{1}\Lambda_{1}^{k-2} & \psi_{2}\Lambda_{2}^{k-2} & \cdots & \psi_{k-1}\Lambda_{k-1}^{k-2} & \phi\zeta^{k-2}\\
\psi_{1}\Lambda_{1}^{n} & \psi_{2}\Lambda_{2}^{n} & \!\! \cdots & \!\! \psi_{n}\Lambda_{n}^{n} & \fbox{0}
\end{array} \right| J_{1}, 
\end{eqnarray}
where $\psi_i(\Lambda_i)$ denote $n$ specified solutions of \eqref{Linear system_NGO} with matrix size $N \times N$. We use the term quasi-Wronskian to call it for short. In fact, since the elements of $J_i$ are noncommutative, the quasiderminant can be considered roughly as a noncommutative version of determinant. 
By the definition\cite{GeRe} , we can decompose \eqref{J_n+1} into 4 blocks as
\begin{eqnarray}
\left|
\begin{array}{cc}
A_{nN \times nN} & B_{nN \times N} \\
C_{N \times nN} & \fbox{$D_{N \times N}$} 
\end{array}
\right|
=
D-CA^{-1}B
\end{eqnarray}
which shows that $J_i$ are all $N \times N$ matrices and $\mathrm{G=GL(N,\mathbb{C})}$. 

If we choose a seed solution $J$ with $\det(J)$ is constant, 
by the Darboux transformation \eqref{Darboux transf}
we have $\det(\widetilde{J})=\det(-\Lambda)\det(J)$ is also a constant.
Applying the Jacobi's formula 
\begin{eqnarray}
\label{Jacobi's formula}
\frac{d}{dt}\mbox{det}A(t) = \mbox{Tr}\left[\mbox{adj}(A(t)) \frac{dA(t)}{dt}\right]
= \mbox{det}A(t)\cdot\mbox{Tr}(A(t)^{-1}\frac{dA(t)}{dt})
\end{eqnarray}
to $\widetilde{J}$ and comparing with \eqref{Gauge fields}, we find that
the gauge fields are all traceless. 
That is, any $J$-matrices \eqref{J_n+1} generated by a seed solution $J_1$ with constant determinant are belonging to $\mathrm{G=SL(N,\mathbb{C})}$ gauge theory.
In fact, we will always set the seed solution $J_1$ to be identity matrix and only consider $N=2, 3$ cases in the next sections.

%Furthermore, if the reader prefer using the Wronskian determinants, one can convert the quasi-Wronskian to the ratios of Wronskians in the following way, for $N=2$ we have 
%\begin{eqnarray}
%\label{1 quasideterminant to 4 quasideterminants}
%\begin{vmatrix}
%A & \begin{array}{cc} B & C \end{array}\\
%\begin{array}{c}
%D \\ E
%\end{array}
%&\fbox{$
%	\begin{array}{cc}
%	f & g \\
%	h & i
%	\end{array}$}
%\end{vmatrix} 
%=
%\left(
%\begin{array}{cc}
%\begin{vmatrix}
%A  & B \\
%D  & \fbox{$f$}
%\end{vmatrix}&
%\begin{vmatrix}
%A & C \\
%D & \fbox{$g$}
%\end{vmatrix}
%\\
%\begin{vmatrix}
%A & B \\
%E & \fbox{$h$}\\
%\end{vmatrix}&
%\begin{vmatrix}
%A & C \\
%E & \fbox{$i$}\\
%\end{vmatrix}
%\end{array}
%\right)  
%=
%\frac{1}{\left| A \right|}
%\left(
%\begin{array}{cc}
%\begin{vmatrix}
%A  & B \\
%D  & f
%\end{vmatrix} &
%\begin{vmatrix}
%A & C \\
%D & g
%\end{vmatrix}
%\\
%\begin{vmatrix}
%A & B \\
%E & h \\
%\end{vmatrix}&
%\begin{vmatrix}
%A & C \\
%E & i \\
%\end{vmatrix}
%\end{array}
%\right), ~~~
%\end{eqnarray} 
%where the size of matrices are 
%$A: k \times k, ~ B, C : k \times 1, ~ D, E : 1 \times k,$ 
%and $f, g, h, i$ are $1 \times 1 ~$ scalar elements.

\section{ASDYM 1-Soliton and Multi-Soliton for $\mathrm{G=SU(2)}$}
Now we set the seed solution $J_1$ to be $2 \times 2$ identity matrix ($\mathrm{G=SL(2,\mathbb{C})}$) so that 
the linear system \eqref{Linear system_NGO} reduces to  
\begin{eqnarray}
\label{Linear system_Reduced}
\left\{
\begin{array}{c}
L(\phi)= \partial_{w}\phi-(\partial_{\widetilde{z}}\phi)\zeta = 0 \\
M(\phi)=	
\partial_{z}\phi-(\partial_{\widetilde{w}}\phi)\zeta = 0   
\end{array}
\right. .
\end{eqnarray}
For the split signature $(+, +, -, -)$, we can take the reality condition \eqref{Reality conidtion_U} and find a lovely solution of \eqref{Linear system_Reduced} as  
\begin{eqnarray}
\begin{array}{l}
\psi
=\left(
\begin{array}{cc}
\!\!ae^{L} &  \!\!\overline{b}e^{-\overline{L}}
\\ 
\!\!-be^{-L} &  \!\! \overline{a}e^{\overline{L}}
\end{array}\right)~ 
\mbox{$w.r.t.$ a specific spectral parameter}~
\Lambda=
\left(
\begin{array}{cc}
\lambda & 0 \\
0 & \overline{\lambda}
\end{array}
\right), ~~~
\end{array}
\end{eqnarray}
where $L=\frac{1}{\sqrt{2}}
\left[
(\lambda\alpha+\beta)x^1
+(\lambda\beta-\alpha)x^2
+(\lambda\alpha-\beta)x^3
+(\lambda\beta+\alpha)x^4 
\right]$, \\
$a$, $b$, $\alpha$, $\beta$, $\lambda \in \mathbb{C}$.

After 1 iteration of the Darboux transformation, we obtain a candidate of 1-soliton solution :
\begin{eqnarray}
\label{J_2}
J_2&=&
\left|
\begin{array}{cc}
\psi & 1 \\
\psi\Lambda & \fbox{0}
\end{array}
\right|
=
-\psi\Lambda \psi^{-1}
  \nonumber  \\
%&&
%L=(\lambda^{(+)} \alpha) z
%+\beta\widetilde{z}
%+(\lambda^{(+)} \beta) w
%+\alpha \widetilde{w},~~\alpha, \beta, z,\widetilde{z}, w, \widetilde{w} \in \mathbb{C}   %\nonumber \\
%&&~~~~~~~\mbox{Split計量 $(+,+,-,-)$ヘのリダクション条件:}   \nonumber  \\
%&&~~~~~~~  
%\left\{
%\begin{array}{l}
%(\lambda^{(+)},\lambda^{(-)}) = (\lambda, \overline{\lambda})
%\\
%\left(\begin{array}{cc}
%\!\!\!z & \!\!\!w \\
%\!\!\!\widetilde{w} & \!\!\!\widetilde{z}
%\end{array}\!\!\right)
%=
%\frac{1}{\sqrt{2}}
%\left(\begin{array}{cc}
%x^{1}+x^{3} & \!\!\!\!x^{2}+x^{4} \\
%\!\!\!\!-(x^{2}-x^{4}) & \!\!\!\!x^{1}-x^{3} 
%\end{array}\!\!\right)
%\end{array}
%\right.	
%\nonumber  \\
&=&
\frac{-1}{\det(\psi)}
\left(
\begin{array}{cc}
\lambda|a|^{2}e^{L+\overline{L}} + \overline{\lambda}|b|^{2}e^{-(L+\overline{L})} &
(\overline{\lambda}-\lambda)a\overline{b}e^{L-\overline{L}}
\\
(\overline{\lambda}-\lambda)\overline{a}be^{-(L-\overline{L})} & \overline{\lambda}|a|^{2}e^{L+\overline{L}}  + \lambda|b|^{2}e^{-(L+\overline{L})}
\end{array}
\right).  ~~~~~
\end{eqnarray}
The resulting action density\cite{HaHu1} of \eqref{J_2} is
\begin{eqnarray}
\begin{array}{l}
{\mbox{Tr}} F_{\mu \nu}F^{\mu \nu}
=
8\left[(\alpha\overline{\beta}-\overline{\alpha}\beta)(\lambda -\overline{\lambda})\right]^2\left(2{\mbox{sech}}^2 X-3{\mbox{sech}}^4 X\right) \\
~~~~~~~~~~~X=L + \overline{L} + \log(\left|a \right| / \left|b \right|)
\end{array}.
\end{eqnarray}
We find that the distribution of action density is localized on a 3-dimensional hyperplane $X=0$. Therefore, this kind of solitons can be interpreted as codimensional 1 soliton and we use the term soliton wall to distinguish them from the domain wall. 

Now we can prepare $n$ different solutions $\psi_i(\Lambda_i)$ of \eqref{Linear system_Reduced} as follows :
\begin{eqnarray}
\begin{array}{l}
\psi_i=\left(
\begin{array}{cc}
a_i e^{L_i} & \overline{b}_ie^{-\overline{L}_i}  \\ 
-b_i e^{-L_i} & \overline{a}_ie^{\overline{L}_i}
\end{array}\!\!\right)~
\mbox{$w.r.t.$ spectral parameters}~
\Lambda_i=
\left(
\begin{array}{cc}
\lambda_i & 0 \\
0 & \overline{\lambda_i}
\end{array}
\right), \\
L_i\!=\!\frac{1}{\sqrt{2}}
\!\left[
(\lambda_i\alpha_i + \beta_i)x^1
\!+\!(\lambda_i\beta_i - \alpha_i)x^2
\!+\!(\lambda_i\alpha_i - \beta_i)x^3
\!+\!(\lambda_i\beta_i + \alpha_i)x^4 
\right], ~~~  \\
~~~~~~~~~~~~a_i, b_i, \alpha_i,\beta_i, \lambda_i \in \mathbb{C},~
i=1,2,\cdots, n.
\end{array}
\end{eqnarray}
After $n$-iterations of the Darboux transformation, we obtain a candidate of $n$-soliton solution : 
\begin{eqnarray}
\label{n-Soliton Solutions_J_n+1}
 J_{n+1} 
 =\left|
 \begin{array}{ccccc}
 \psi_{1} & \psi_{2} &  \cdots &  \psi_{n} & 1 \\
 \psi_{1}\Lambda_{1} & \psi_{2}\Lambda_{2} &  \cdots & \psi_{n}\Lambda_{n} & 0 \\
 \vdots &  \vdots &  \ddots &  \vdots & \vdots  \\
 %\psi_{1}\Lambda_{1}^{k-2} & \psi_{2}\Lambda_{2}^{k-2} & \cdots & \psi_{k-1}\Lambda_{k-1}^{k-2} & \phi\zeta^{k-2}\\
 \psi_{1}\Lambda_{1}^{n} & \psi_{2}\Lambda_{2}^{n} &  \cdots & \psi_{n}\Lambda_{n}^{n} & \fbox{0}
 \end{array} \right|, 
\end{eqnarray}
which satisfies the property\cite{HaHu2, SCH}
\begin{eqnarray}
\label{J_n+1 J_n+1^{dagger}}
J_{n+1}J_{n+1}^{\dagger}=J_{n+1}^{\dagger}J_{n+1}=\prod_{i=1}^{n}|\lambda_i|^2I,~~ I : 2\times 2~ \mbox{identity matrix}.
\end{eqnarray}
This fact implies that the gauge fields in 4-dimensional split signature
\begin{eqnarray}
\label{Real representation of gauge fields_U_muti-soliton} 
\begin{array}{l}
A_{1}^{(n+1)} = A_{3}^{(n+1)}
=\frac{-1}{2}
\left[(\partial_{1}J_{n+1})J_{n+1}^{-1}+(\partial_{3}J_{n+1})J_{n+1}^{-1} \right]   \\
A_{2}^{(n+1)}=A_{4}^{(n+1)}
=\frac{-1}{2}
\left[(\partial_{2}J_{n+1})J_{n+1}^{-1} + (\partial_{4}J_{n+1})J_{n+1}^{-1} \right]  
\end{array}
\end{eqnarray}
are all anti-hermitian and therefore the gauge group is $\mathrm{G=SU(2)}$.

Now a natural question to ask is whether the $n$-soliton solution \eqref{n-Soliton Solutions_J_n+1} 
gives rise to $n$ intersecting soliton walls or not.
Let us use a quite similar technique as mentioned in reference \refcite{MaSa} to
discuss the asymptotic action density of \eqref{n-Soliton Solutions_J_n+1} rather than calculating the action density directly by \eqref{Real representation of gauge fields_U_muti-soliton}. 
First of all, we fix an $I \in \left\{1, 2,..., n \right\}$ and consider a comoving frame related to the $I$-th 1-soliton solution :
\begin{eqnarray}
\label{I-th 1-soliton}
{J}_{2}^{~\!(I)}
=
-\psi_{n}^{(I)}\Lambda_{I}(\psi_{n}^{(I)})^{-1},  ~~   
\psi_n^{(I)}=\left(
\begin{array}{cc}
a_I e^{L_I} & \overline{b}_I ~\!e^{-\overline{L}_I} \\
-b_I e^{-L_I} & \overline{a}_I ~\! e^{\overline{L}_I}
\end{array}
\right)
\end{eqnarray}
whose action density is 
\begin{eqnarray}
\label{Action density_I-th 1-soliton}
\begin{array}{l}
{\mbox{Tr}} F_{\mu \nu}F^{\mu \nu ~\!(I)}
=
8\left[(\alpha_I\overline{\beta}_I-\overline{\alpha}_I\beta_I)
(\lambda_I -\overline{\lambda}_I)  \right]^2
\left(2{\mbox{sech}}^2 X_I-3{\mbox{sech}}^4 X_I\right) \\
~~~~~~~~~~~~~X_I=L_I + \overline{L}_I + \log(\left|a_I \right| / \left|b_I \right|) 
\end{array}. ~~~~~
\end{eqnarray}
\smallskip \\
More precisely, we define $r:=\sqrt{(x_1)^{2}+(x_2)^{2}+(x_3)^{2}+(x_4)^{2}}$  and consider the asymptotic limit $r \rightarrow \infty$ such that
\begin{eqnarray}
\left\{
\begin{array}{l}
X_{I} ~\mbox{is a finite real number}  \\
X_{i, i \neq I} ~\rightarrow  \pm \infty  ~~(i.e. ~{\mbox{Tr}} F_{\mu \nu}F^{\mu \nu (i \neq I)} \rightarrow 0)
\end{array}.
\right. 
\end{eqnarray}
By some mathematical techniques\cite{HaHu2, SCH} of the quasideterminant, we find that
\begin{eqnarray}
\label{Asymptotic form of J_{n+1}_psi_I}
J_{n+1} 
&~\stackrel{r \rightarrow \infty}{\longrightarrow}~&
- \widetilde{\Psi}_{n}^{~\!(I)}\Lambda_{I} (\widetilde{\Psi}_{n}^{~\!(I)})^{-1} D_{n}^{~\!(I)}, ~n \geq 2, ~~~~~ 
\end{eqnarray}
where $D_n^{~\!(I)}$ is a constant matrix and doesn't affect the gauge fields, and
\begin{eqnarray}
\label{tilde{Psi}_n}
\widetilde{\Psi}_{n}^{~\!(I)}:=
\left\{
\begin{array}{l}
(i)~\!
\left(
\begin{array}{cc}
\prod\limits_{i=1,i \neq I}^{n}(\lambda_{I} - \lambda_{i})~\! a_Ie^{L_I}
&
\prod\limits_{i=1,i \neq I}^{n}(\overline{\lambda}_{I} - \lambda_{i})~\!\overline{b}_I~\! e^{-\overline{L}_I}\!
\\
-\!\!\!\prod\limits_{i=1,i \neq I}^{n}(\lambda_{I} -\overline{\lambda}_{i})~\!b_{I}e^{-L_I}
&
\!\!\!\prod\limits_{i=1,i \neq I}^{n}(\overline{\lambda}_{I} - \overline{\lambda}_{i})~\!\overline{a}_I~\! e^{\overline{L}_I}\!
\end{array}
\right)
\\
~~~~~~~~\mbox{as $X_{i, i \neq I} \rightarrow + \infty$}
\\
(ii)
\left(
\begin{array}{cc}
\prod\limits_{i=1,i \neq I}^{n}(\lambda_{I} - \overline{\lambda}_{i})~\! a_Ie^{L_I}
&
\prod\limits_{i=1,i \neq I}^{n}(\overline{\lambda}_{I} - \overline{\lambda}_{i})~\!\overline{b}_I~\! e^{-\overline{L}_I}\!
\\
-\!\!\!\prod\limits_{i=1,i \neq I}^{n}(\lambda_{I} -\lambda_{i})~\!b_{I}e^{-L_I}
&
\!\!\!\prod\limits_{i=1,i \neq I}^{n}(\overline{\lambda}_{I} - \lambda_{i})~\!\overline{a}_I~\! e^{\overline{L}_I}\!
\end{array}
\right)
\\
~~~~~~~~\mbox{as $X_{i, i \neq I} \rightarrow -\infty$}
\end{array}
\right.. ~~~~~~~
\end{eqnarray}
Comparing the asymptotic $n$-solion solution \eqref{Asymptotic form of J_{n+1}_psi_I}, \eqref{tilde{Psi}_n} with $I$-th 1-soliton solution \eqref{I-th 1-soliton},
we find that $\widetilde{\Psi}_n^{~\!(I)}$ is in the same form as $\psi_{n}^{(I)}$ up to constant factors. These factors lead to a position shift from the principal peak of action density \eqref{Action density_I-th 1-soliton}, called phase shift. 
More specifically, the action density of $n$-soliton solution in the asymptotic region
\begin{eqnarray}
	{\mbox{Tr}} F_{\mu \nu}F^{\mu \nu}
	\stackrel{r \rightarrow \infty}{\longrightarrow}
	8\left[(\alpha_I\overline{\beta}_I-\overline{\alpha}_I\beta_I)
	(\lambda_I -\overline{\lambda}_I)  \right]^2
	\left(2{\mbox{sech}}^2 \widetilde{X}_I-3{\mbox{sech}}^4 \widetilde{X}_I\right)~~~~ 
\end{eqnarray}
behaves like the action density of $I$-th 1-soliton \eqref{Action density_I-th 1-soliton},
where $\widetilde{X}_I:=X_I + \Delta_{I}$ and the phase shift
\begin{eqnarray}
\label{Phase shift_G=SU(2)_U}
\Delta_{I}
=
\sum_{i=1,i\neq I}^{n}\varepsilon_i^{(\pm)}
\log 
\left|
\frac{\lambda_I-\lambda_i}{\lambda_I-\overline{\lambda}_i} 
\right|,~~
\left\{
\begin{array}{l}
\varepsilon_i^{(+)}:=+1, ~~ X_{i,i \neq I} \rightarrow +\infty \\
\varepsilon_i^{(-)}:=-1, ~~ X_{i,i \neq I} \rightarrow -\infty
\end{array}
\right.
\end{eqnarray}
is real-valued. Since $I$ is any positive integer from 1 to $n$, and for every $I$ the $n$-soliton solution gives rise to a soliton wall in the asymptotic region, we can conclude that  the $n$-soliton solution can be interpreted as $n$ intersecting soliton walls in the entire region. Furthermore, the $n$ intersecting soliton walls can be embedded into $\mathrm{G=SU(2)}$ gauge theory in 4-dimensional split signature and therefore they could be interpreted as $n$ intersecting branes in open $\mathrm{N=2}$ string theories.

\section{Reduction to (1+1)-dimensional real space}

To make the discussion more clearly, let us take 2-dimensional spacetime and 3-soliton scattering for instance.
We can impose the condition $x^{1}=t$, $x^{2}=x^{4}=0$, $x^{3}=x$ on the spacetime coordinates such that
\begin{eqnarray}
L_i=
\frac{1}{\sqrt{2}}
\left(
(\lambda_i\alpha_i+\beta_i)t
%+(\lambda_i\beta_i-\alpha_i)x^2
+(\lambda_i\alpha_i-\beta_i)x
%+(\lambda_i\beta_i+\alpha_i)x^4 
\right),~i=1,2,3
\end{eqnarray}
and $X_{i}=L_i + \overline{L}_i +\log\left|a_i/b_i\right|$. 
Then we fix an $I \in \left\{1,2,3 \right\}$ and choose a complex number $\ell_{I}$ such that $L_{I}=\ell_{I}$ which implies 
\begin{eqnarray}
L_i=
\left(\frac{\lambda_i\alpha_i-\beta_i}{\lambda_I\alpha_I-\beta_I}\right)\ell_I
+\sqrt{2}\left(
\frac{\lambda_i\alpha_i\beta_I-\lambda_I\alpha_I\beta_i}{\lambda_I\alpha_I-\beta_I}
\right)t, ~~i \neq I.
\end{eqnarray}
Now the setup of the comoving frame related to the $I$-th 1-soliton is completed because
%Now we complete the setup of the comoving frame related to the $I$-th 1-soliton yet because
\begin{eqnarray}
\left\{
\begin{array}{l}
X_{I} = \mbox{finite} \\
X_{i, i \neq I} \rightarrow \pm \infty ~~ \mbox{or} ~\mp \infty 
\end{array}
\right.
~~\mbox{when}~~  t \rightarrow \pm \infty
\end{eqnarray}
This setup implies that the action density of $i$-th 1-soliton ($i \neq I$) behaves as  Tr$F_{\mu\nu}F^{\mu\nu ~\!(i \neq I)} \sim 2\mbox{sech}^2X_{i}-3\mbox{sech}^4X_{i} \rightarrow 0$ as $t \rightarrow \pm \infty$.
On the other hand, the action density of 3-soliton in the asymptotic region behaves as 
\begin{eqnarray}
\label{X_I=Delta_2D}
\begin{array}{l}
\mbox{Tr}F_{\mu\nu}F^{\mu\nu}  \sim   2\mbox{sech}^2\widetilde{X}_{I}-3\mbox{sech}^4\widetilde{X}_{I}, ~
\widetilde{X}_{I}    
=\!
\left\{
\begin{array}{l}
X_{I} + \Delta_{I}^{(+)}  ~ \mbox{ as } ~ t \rightarrow +\infty \\ 
X_{I} + \Delta_{I}^{(-)}  ~ \mbox{ as } ~ t \rightarrow -\infty
\end{array}.
\right. ~~~ 
\end{array}
\end{eqnarray}
In fact, $\Delta_{I}^{(-)}=-\Delta_{I}^{(+)}$ and it depends on $2^{3-1}=4$ choices of asymptotic regions.
For example one of the choices is 
\begin{eqnarray}
\Delta_{1}^{(+)}
&=&
-\Delta_{1}^{(-)}
=
-\log\left| \frac{\lambda_1-\lambda_2}{\lambda_1-\overline{\lambda}_2} \right|
-\log\left| \frac{\lambda_1-\lambda_3}{\lambda_1-\overline{\lambda}_3} \right|,~~~~   \\
\Delta_{2}^{(+)}
&=&
-\Delta_{2}^{(-)}
=
+\log\left| \frac{\lambda_2-\lambda_1}{\lambda_2-\overline{\lambda}_1} \right|
-\log\left| \frac{\lambda_2-\lambda_3}{\lambda_3-\overline{\lambda}_3} \right|,~~~~ \\
\Delta_{3}^{(+)}
&\!\!\!\!=\!\!\!\!&
-\Delta_{3}^{(-)}
=
+\log\left| \frac{\lambda_3-\lambda_1}{\lambda_3-\overline{\lambda}_1} \right|
+\log\left| \frac{\lambda_3-\lambda_2}{\lambda_3-\overline{\lambda}_2} \right|.~~~~ 
\end{eqnarray}

\section{An example of ASDYM 1-Soliton for $\mathrm{G=SU(3)}$ and Multi-Soliton scattering }
 
If we consider the linear system \eqref{Linear system_Reduced} for $\mathrm{G=SL(3,\mathbb{C})}$, 
these is a special class of solutions $\psi_i(\Lambda_i)$ that give rise to soliton walls \cite{SCH} and behave like the $\mathrm{G=SU(2)}$ cases as well. Firstly, the candidate of $\mathrm{G=SL(3,\mathbb{C})}$ 1-soliton solution can be constructed by 1-iteration of the Darboux transformation
\begin{eqnarray}
\label{1-Soliton solution_U_G=SU(3)}
\begin{array}{l}
J_2^{~\!(i)}
=
-\psi_i\Lambda_i \psi_i^{-1}, ~~  
\left\{
\begin{array}{l}
\psi_i
:=\left(
\begin{array}{ccc}
a_ie^{L_i}
& 
\overline{b}_ie^{-\overline{L}_i}
&
0
\\ 
-b_ie^{-L_i}
& 
\overline{a}_ie^{\overline{L}_i}  
&
c_ie^{-L_i}
\\
0
&
-\overline{c}_ie^{-\overline{L}_i}
&
a_ie^{L_i}
\end{array}
\right),~ \\
\Lambda_i:=
\left(
\begin{array}{ccc}
\lambda_i & 0 & 0\\
0 & \overline{\lambda}_i & 0 \\
0 & 0 & \lambda_i
\end{array}
\right)
\end{array},
\right.   
\\
L_i=\frac{1}{\sqrt{2}}
\left[
(\lambda_i\alpha_i + \beta_i)x^1
+(\lambda_i \beta_i -\alpha_i)x^2
+(\lambda_i \alpha_i -\beta_i)x^3
+(\lambda_i \beta_i + \alpha_i)x^4 
\right],    \\
~~~~~~~~~~~~ a_i, b_i, c_i, \alpha_i, \beta_i, \lambda_i \in \mathbb{C},~ i = 1,2,...,n. 
\end{array}
\end{eqnarray}
By direct calculation, we obtain the action density of $i$-th 1-soliton as 
\begin{eqnarray}
\label{Reduced Lagrangian density_U_G=SU(3)}
\begin{array}{l}
{\mbox{Tr}} F_{\mu \nu}F^{\mu \nu ~\!(i)}
=
8\left[(\alpha_i\overline{\beta}_i -\overline{\alpha}_i\beta_i)
(\lambda_i -\overline{\lambda}_i)  \right]^2
\left(2{\mbox{sech}}^2 X_i -3{\mbox{sech}}^4 X_i\right) \\
X_i = L_i + \overline{L}_i + \frac{1}{2}\log\left[\left|a_i \right|^2 / (\left|b_i \right|^2 +\left|c_i \right|^2)\right]
\end{array}, ~~~~
\end{eqnarray}
which can be interpreted as $i$-th soliton wall as discussed in $\mathrm{G=SU(2)}$ case.
Substituting $\psi_i$ of \eqref{1-Soliton solution_U_G=SU(3)} into \eqref{n-Soliton Solutions_J_n+1}, we get a version of $n$-soliton $J_{n+1}$ for $\mathrm{G=SL(3,\mathbb{C})}$. 
By some mathematical techniques\cite{SCH} of the quasideterminant and considering a comoving frame related to the $I$-th 1-soliton, we find that
\begin{eqnarray}
\label{Asymptotic form of J_{n+1}_psi_I_SU(3)}
J_{n+1} 
\stackrel{r \rightarrow \infty}{\longrightarrow}
\widetilde{J}_{n+1}^{~\!(I)}
=:
- \widetilde{\Psi}_{n}^{~\!(I)}\Lambda_{I} (\widetilde{\Psi}_{n}^{~\!(I)})^{-1} D_{n}^{~\!(I)}, ~n \geq 2, ~~~~~ 
\end{eqnarray}
where $D_n^{~\!(I)}$ is a constant matrix and $\widetilde{\Psi}_{n}^{~\!(I)}$ is in the form of 
\begin{eqnarray}
\label{Qn_I_reduced form_G=SU(3)}
\widetilde{\Psi}_{n}^{~\!(I)}
=
\left(
\begin{array}{ccc}
A_Ie^{L_I} & \overline{B}_Ie^{-\overline{L}_I} & 0 \\
-B_Ie^{-L_I} & \overline{A}_Ie^{\overline{L}_I} & C_I e^{-L_I} \\
0 & -\overline{C}_Ie^{-\overline{L}_I} & A_Ie^{L_I}
\end{array}
\right),~
\left\{
\begin{array}{l}
A_I=\!\!\!\!\prod\limits_{i=1,i \neq I}^{n}(\lambda_{I}-\lambda_{i}^{(\pm)})~\!a_I
\\
B_I=\!\!\!\!\prod\limits_{i=1,i \neq I}^{n}(\lambda_{I}-\lambda_{i}^{(\mp)})~\!b_{I}
\\
C_I=\!\!\!\!\prod\limits_{i=1,i \neq I}^{n}(\lambda_{I}-\lambda_{i}^{(\mp)})~\!c_{I}
\\
(\lambda_i^{(+)}, \lambda_i^{(-)}):=(\lambda_i, \overline{\lambda}_i)
\end{array}
\right. ~~~
\end{eqnarray}
Comparing \eqref{Qn_I_reduced form_G=SU(3)} with \eqref{1-Soliton solution_U_G=SU(3)} and \eqref{Reduced Lagrangian density_U_G=SU(3)}, we can conclude that the action density of $n$-soliton in the asymptotic region
\begin{eqnarray}
{\mbox{Tr}} F_{\mu \nu}F^{\mu \nu}
\stackrel{r \rightarrow \infty}{\longrightarrow}
8\left[(\alpha_I\overline{\beta}_I-\overline{\alpha}_I\beta_I)
(\lambda_I -\overline{\lambda}_I)  \right]^2
\left(2{\mbox{sech}}^2 \widetilde{X}_I-3{\mbox{sech}}^4 \widetilde{X}_I \right) ~~~
\end{eqnarray} 
behaves like the action density of $I$-th 1-soliton,
where $\widetilde{X}_I:=X_I + \Delta_{I}$ and the phase shift is
\begin{eqnarray}
\Delta_{I}
=
\frac{1}{2}
\log 
\left[
\frac{\prod\limits_{i=1, i \neq I}^{n}\left|\lambda_I-\lambda_i^{(\pm)}\right|^2(\left| b_I\right|^2 + \left| c_I\right|^2)}
{\prod\limits_{i=1, i \neq I}^{n}\left|(\lambda_I-{\lambda}_i^{(\mp)})\right|^2\left|b_I\right|^2 +\prod\limits_{i=1, i \neq I}^{n}\left|(\lambda_I-\lambda_i^{(\mp)})\right|^2\left|c_I\right|^2} 
\right] ~~~~
\end{eqnarray}
which is real-valued and depends on $2^{n-1}$ choices of $(\lambda_i^{(+)}, \lambda_i^{(-)}):=(\lambda_i, \overline{\lambda}_i)$.
Therefore, the $n$-soliton can be interpreted as $n$-intersecting soliton walls for $\mathrm{G=SL(3,\mathbb{C})}$. 
On the other hand, the asymptotic form $\widetilde{J}_{n+1}^{~\!(I)}$ of the $n$-soliton solution $J_{n+1}$ satisfies the property\cite{SCH}
\begin{eqnarray}
\widetilde{J}_{n+1}^{~\!(I)~\!\dagger}\widetilde{J}_{n+1}^{~\!(I)}
=\widetilde{J}_{n+1}^{~\!(I)}\widetilde{J}_{n+1}^{~\!(I)~\!\dagger}
=\prod_{i=1}^{n}\left| \lambda_i \right|^2 I_{3 \times 3}
\end{eqnarray}
which implies that the gauge fields given by $\widetilde{J}_{n+1}^{~\!(I)}$ are all anti-hermitian.
Therefore for each single soliton wall in the asymptotic region, the gauge group is in fact $\mathrm{G=SU(3)}$. 

\section{Conclusion}
In this paper, we found a class of ASDYM 1-solitons in 4-dimensional split signature for $\mathrm{G=SU(2)}$ and $\mathrm{SU(3)}$, respectively. The resulting action densities are in the same form as
${\mbox{Tr}} F_{\mu \nu}F^{\mu \nu} \propto \left(2{\mbox{sech}}^2 X-3{\mbox{sech}}^4 X\right)$
which can be interpreted as the soliton walls.
After $n$-iterations of the Darboux transformation, we obtain the ASDYM $n$-soliton and interpret it as $n$ intersecting soliton walls for $\mathrm{G=SU(2)}$ and $\mathrm{SL(3,\mathbb{C})}$, respectively.
This fact is a well-known feature for the KdV multi-solitons, but a new insight for 
the ASDYM multi-solitons.  
On the other hand, the $n$ intersecting soliton walls can be embedded into $\mathrm{G=SU(2)}$ gauge theory  
and hence they could be interpreted as $n$ intersecting branes in $\mathrm{N=2}$ open string theories.  
Therefore, to understand the role of such physical objects that play in $\mathrm{N=2}$ open string theories would be an interesting future work, and the relationship between $\mathrm{N=2}$ open string theories and lower-dimensional integrable systems is also worth studying.

\end{document}